\documentstyle[11pt,newpasp,twoside,epsf]{article}
\def\edcomment#1{\iffalse\marginpar{\raggedright\sl#1\/}\else\relax\fi}
\reversemarginpar

\begin{document}
\title{The Large-Scale Environment of Dynamical Young Clusters of Galaxies}
 \author{Manolis Plionis}
\affil{Institute of Astronomy \& Astrophysics, National Observatory
of Athens, I.Metaxa \& B.Pavlou, P.Penteli, 152 36 Athens, Greece}

\author{Spyros Basilakos}
\affil{Astrophysics Group, Blackett Laboratory, Imperial College,
London, UK}

\begin{abstract}
Using the APM cluster data (Dalton et al. 1997) 
we investigate whether the dynamical status of
clusters is related to the large-scale structure of the Universe. The
methods used have been calibrated using a limited number of APM/Abell clusters
for which ROSAT pointed observations were available.
We find that cluster substructure is strongly correlated with the
tendency of clusters to be aligned with their nearest neighbour and in
general with the nearby clusters that belong to the same
supercluster. Furthermore, dynamically young clusters are more 
clustered than the overall cluster population.
These are strong indications that clusters develop in a
hierarchical fashion by merging along the 
large-scale filamentary structures within
which they are embedded.
\end{abstract}

\section{Introduction}
In the popular hierarchical clustering scenario, clusters are supposed
to form by merging and accretion of galaxies along filamentary
protostructures. This structure formation scenario is seen to work
in cosmological N-body CDM simulations, while observational studies
have provided indications that our Universe
follows such a scenario 
(cf. West 1994; West, Jones \& Forman 1995)

If we can unambiguously identify the dynamical status of clusters
(eg. presence of significant substructures, X-ray
temperature gradients, compression of X-ray emitting gas along the merging
direction, cf. Sarazin 2000; Schindler 1999) 
then we can search for indications of cluster
formation along the lines of the hierarchical scenario.

Mohr et al. (1995), Rizza et al. (1998)
and Kolokotronis et al. (2001) have investigated 
the frequency of cluster substructure using in a 
complementary fashion optical and X--ray data. The 
advantage of using X--ray data is that the X--ray emission is proportional 
to the square of the gas density (rather than just density in the optical) 
and therefore it is centrally concentrated, a fact which minimises 
projection effects. The advantage of 
using optical data is the shear size of the available cluster 
catalogues and thus the statistical significance of the emanating results. 
Kolokotronis et al. (2001) calibrated various substructure measures
using APM and ROSAT data of 22 Abell clusters
and found that in most cases using X--ray or optical data one can
identify substructure unambiguously. Only in $\sim 20\%$ of the
clusters they studied did they find projection effects in
the optical that altered the X-ray definition of substructure. Their
conclusion was that solely optical cluster imaging data can be used 
in order to identify the clusters that have significant substructure.

An interesting observable, that was thought initially to provide 
strong constraints on theories of galaxy formation, 
is the tendency of clusters to be aligned with their nearest 
neighbour as well as with other clusters that reside in the same 
supercluster (cf. Binggeli 1982; Plionis 1994). 
Analytical (Bond 1986) and numerical
(cf. Splinter et al. 1997; Onuora \& Thomas 2000)
work have shown that such alignments, expected naturally to occur in 
"top-down" scenarios, are also found in hierarchical clustering models of 
structure formation like the CDM. 

\section{Substructure \& Alignment Measures}
Evrard et al. (1993) and Mohr et al. (1995)
have suggested as an 
indicator of cluster substructure the shift of the center-of-mass 
position as a function of density threshold above which 
it is estimated. The {\em centroid-shift} ($sc$) is defined
as the distance between the cluster 
center-of-mass, which is a function of density threshold, and the
highest cluster density-peak.
The significance, $\sigma$, of such centroid variations to
the presence of background contamination and random density
fluctuations, are estimated using 
Monte Carlo cluster simulations in which, by construction, there is no
substructure (see details in Plionis 2001).
According to Kolokotronis et al. (2001), a large and
significant value of $sc$ is a clear indication of substructure in the
optical APM cluster data.

Furthermore, 
in order to investigate the alignment between cluster orientations,
we use the relative position angle
between cluster pairs, $\phi_{i,j}$.
In an isotropic distribution we will have
$\langle \phi_{i,j} \rangle \simeq 45^{\circ}$. 
A significant deviation from this would be an indication of an 
anisotropic distribution which can be quantified by 
$\delta=\sum \phi_{i,j}/N-45$ (Struble \& Peebles 1995).
In an isotropic distribution we have $\langle \delta \rangle \simeq 0$, while
the standard deviation is given by $\sigma=90/\sqrt{12 N}$. 
A significantly negative value of $\delta$ would indicate alignment and 
a positive misalignment.

\section{Results: Alignments vs Substructure:}
We have correlated the alignment signal with the substructure
significance indication, $\sigma$, in order to see whether there is any relation
between the large-scale environment, in which the cluster distribution
is embedded, and the internal cluster dynamics. 
In the left panel of figure 1 we present the alignment signal, 
$\langle \delta \rangle$,
between cluster nearest-neighbours (filled dots) and between all pairs
(open dots) with pair separations $< 20$ $h^{-1}$ Mpc. 
Evidently, there is a strong correlation
between the strength of the alignment signal and the substructure
significance level. 
\begin{figure}
\plottwo{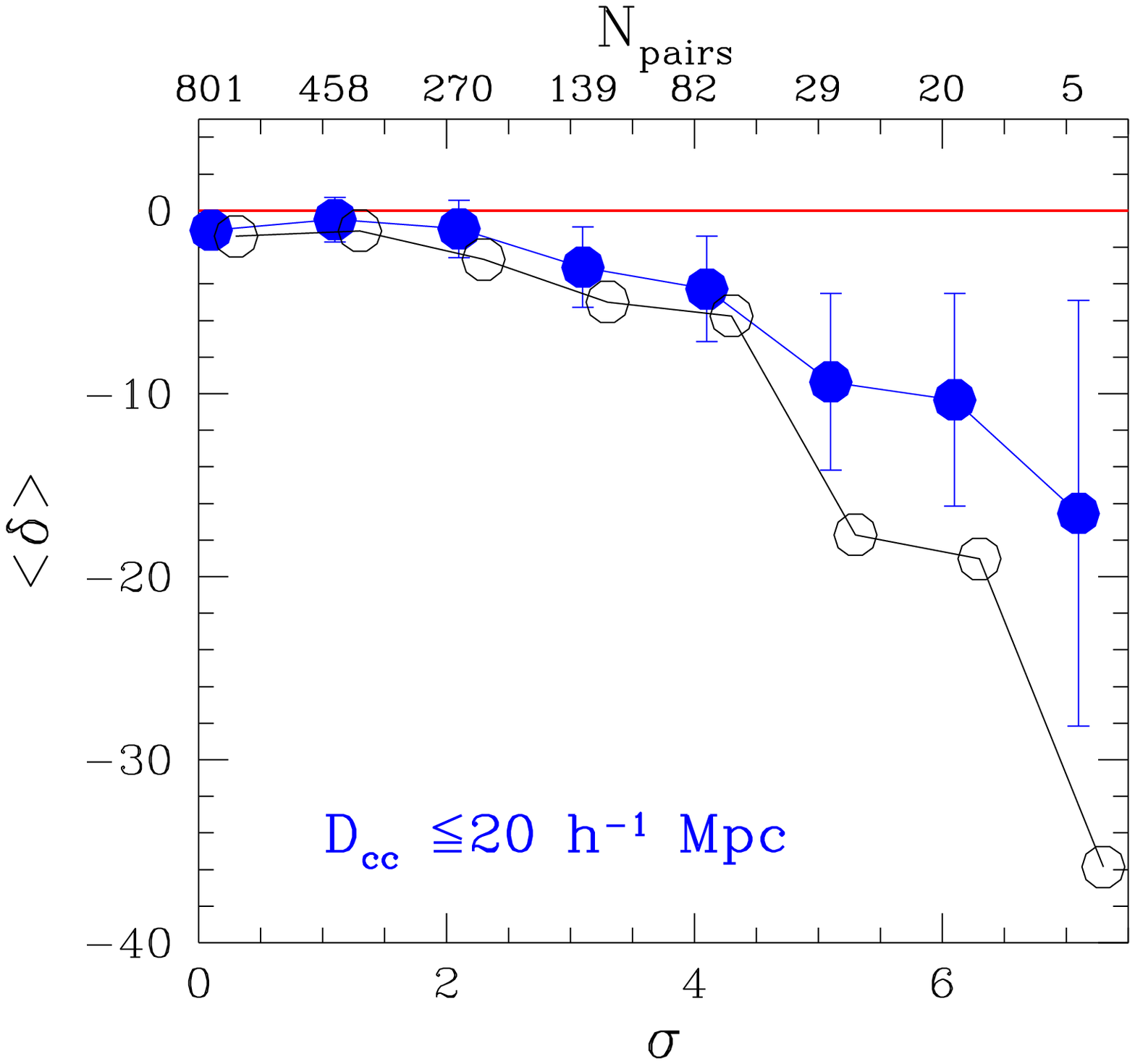}{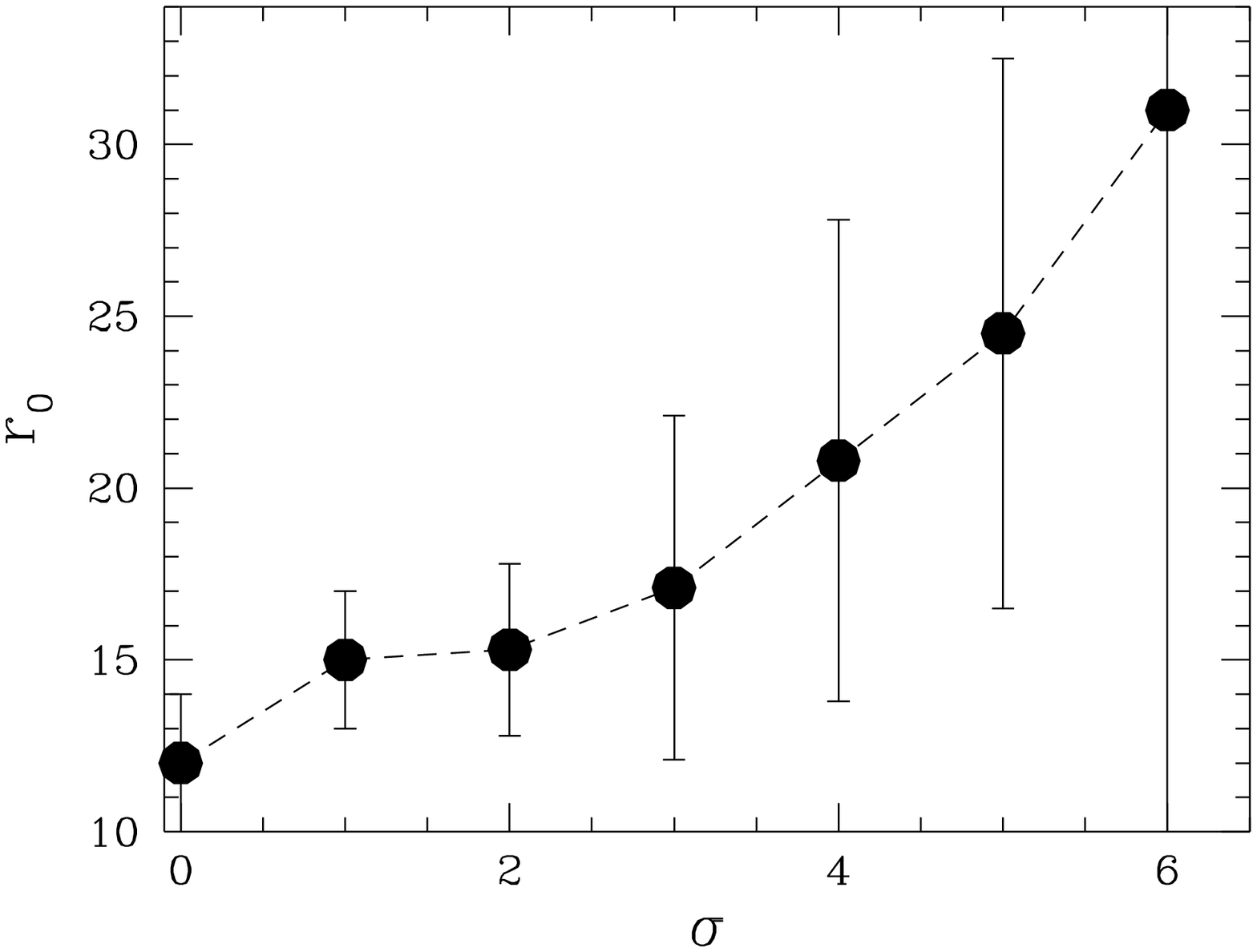}
\caption{Left: Alignment signal as a function of substructure 
significance level. 
Filled circles correspond to nearest-neighbours and open circles 
to all neighbours within 20 $h^{-1}$ Mpc. 
Right: The APM cluster correlation length as a function of 
substructure significance level.}
\end{figure}

\section{Results: Local Density vs Substructure}

Where do clusters with significant substructure reside?
To answer this question we measure the
spatial 2-point cluster correlation function for different
substructure significance levels.
In the right panel of figure 1 we plot the correlation length, $r_{0}$, as a
function of cluster substructure significance level, which is 
clearly an increasing function of $\sigma$.
The conclusion
of this analysis is that clusters
showing evidence of dynamical activity reside in high-density
environments. Our results are in agreement with a similar 
environmental dependence found in
the REFLEX and BCS X-ray cluster sample (Sch\"{u}ecker et al., this volume) and
for cooling flow clusters with high mass accretion rates (Loken, Melott
\& Miller 1999).

\begin{figure}
\center
\begin{minipage}[b]{0.69\linewidth}
\plotone{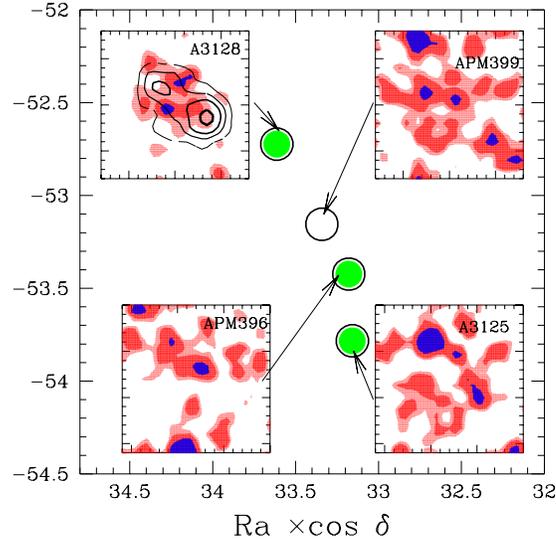}
\end{minipage}
\caption{An example of a dense APM supercluster. 
Filled dots signify clusters with
significant substructure. For A3128 we overlay the ROSAT PSPC
X-ray contours.}
\end{figure}
 
As an individual 
example we present in figure 3 the distribution on the sky of a
dense supercluster, found with a percolation radius of 5 $h^{-1}$ Mpc.
Embedded in the plot are the galaxy density distributions of the
clusters that constitute the supercluster. It is interesting that
the position angles of the clusters (between
16$^{\circ}$ and 48$^{\circ}$) are aligned also with the
orientation of the filamentary supercluster, which has a position angle
of $\sim 30^{\circ}$.

\end{document}